\documentclass[fleqn,twoside]{article}
\usepackage{espcrc2}

\readRCS
$Id: espcrc2.tex,v 1.2 2004/02/24 11:22:11 spepping Exp $
\usepackage{graphicx}

\def\z0{\rm Z^0}
\def\as{\alpha_{\rm s}}

\newcommand{\epem}{{\rm e^+\rm e^-}}

\newcommand{\amz}{\as(M_{\rm Z^0})}

\def\mz{M_{\rm Z^0}}

\def\rz{R_{\rm Z}}
\def\d2{D_2}
\def\oq{\char'134}
\def\lamsb{\Lambda_{\overline{\mbox{\scriptsize MS}}}}

\def\m2{\mu^2}
\def\q{\rm q}

\def\p{\rm p}

\def\q2{Q^2}
\def\asq{\as (\q2 )}
\def\wamz{\overline{\as}(M_{\rm Z^0})}

\def\dwas{\Delta\overline{\as}}

\def\r3{R_3}

\def\etj{E^{jet}_T}
\newcommand{\AmS}{{\protect\the\textfont2
  A\kern-.1667em\lower.5ex\hbox{M}\kern-.125emS}}

\title{$\as$ at Zinnowitz 2004}

\author{Siegfried Bethke\address{Max Planck Institute of Physics, \\ 
        F\"ohringer Ring 6, 80805 Munich, Germany}}
       
\begin{document}

\begin{abstract}
A review of measurements of $\as$ is given, representing the status of April 2004.
The results prove the energy dependence of $\as$ and are in 
excellent agreement with the expectations of Quantum Chromodynamics, QCD.
Evolving all results to the rest energy of the $\z0$ boson, 
the world average of 
$\amz$ is determined from measurements which are based on QCD calculations in complete NNLO perturbation theory, giving
$$ \amz = 0.1182 \pm 0.0027 \ . $$
\vspace{1pc}
%
%

\end{abstract}

\maketitle

\section{INTRODUCTION}

The coupling constant of the Strong Interactions, $\as$, is 
one of the most fundamental parameters of nature which is to be determined 
by experiment.
In this review, a summary
of the most recent measurements of $\as$ 
representing the status of April 2004
is given, providing 
another incremental update of a more complete and concise review \cite{concise} and of \cite{as2002}.
For a detailed introduction into the field and for an overview and 
definition of basic concepts, equations and 
references, the reader is referred to~\cite{concise,as2002}.

\section{NEW RESULTS}

New or updated measurements of $\as$ are available from many 
classes of high energy particle reactions. 
In the following subsections, the respective results will be shortly reviewed.

\subsection{Deep Inelastic Scattering (DIS)}

New measurements of $\as$ from inclusive jet cross sections in $\gamma$p interactions at HERA 
are available from the ZEUS collaboration \cite{zeus03}.
Jet cross sections and values of $\as$ are presented as a function
of the jet transverse energy, $\etj$, for jets with
$\etj > 17$~GeV.
The resulting values of $\as (\etj )$, based on NLO QCD calculations,
are displayed in Figure~1.
They are in good agreement with the running of $\as$, as expected by QCD,
and average to
\begin{eqnarray}
\amz = 0.1224 &\pm 0.0001&{\rm (stat.)} \nonumber \\
&^{+0.0022}_{-0.0019}&{\rm(exp.)} \nonumber \\
&^{+0.0054}_{-0.0042}&{\rm(theo.)},
\end{eqnarray}
if evolved to the energy scale of $\mz$ using the QCD $\beta$ function 
in two-loop approximation, c.f. \cite{concise} Equation~7.

Averaging all measurements of $\as$ from jet production at HERA, which 
were recently reviewed e.g. in \cite{buschhorn}, results in
\begin{eqnarray}
\amz = 0.120 &\pm 0.002& {\rm (exp.)} \nonumber \\
&\pm 0.004& {\rm (theo.)},
\end{eqnarray}
unchanged from the previous average \cite{as2002}.

\begin{figure}[htb]
\vskip-10mm
\begin{center}
\includegraphics[width=73mm, keepaspectratio]{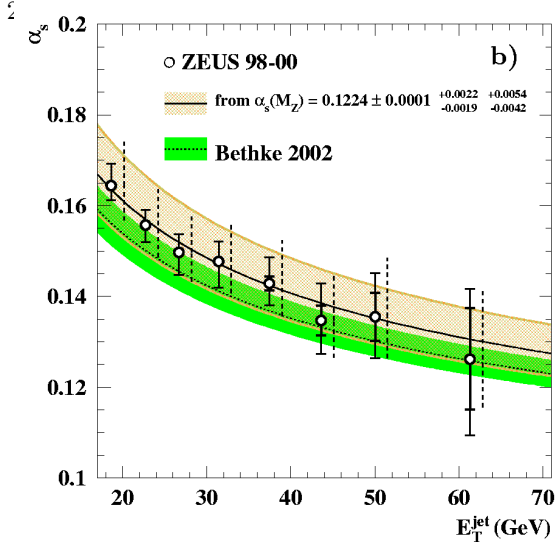}
\end{center}
\caption{\label{zeus2}
$\as (\etj )$ values determined from QCD fits of the measured
$d \sigma \over d\etj$ from ZEUS \cite{zeus03}.
The solid line corresponds to the central value of $\amz$ determined in
\cite{zeus03}, the dashed line represents the world average \cite{as2002}.
}
\end{figure}

An update of $\as$ from measurements of
polarized structure functions \cite{bluemlein02}, 
now containing
data from the HERMES experiment at the HERA collider, results in
\begin{eqnarray}
\amz = 0.113 &\pm 0.004& {\rm (exp.)} \nonumber \\
&^{+0.009}_{-0.006}& {\rm (theo.)},
\end{eqnarray}
in NLO QCD.

A new global analysis using all available precision data of deep inelastic
and related hard scattering processes includes recent measurements of
structure functions from HERA and of the inclusive jet cross sections at the
Tevatron \cite{mrst03}.
After analysis of experimental and theororetica uncertainties, and restrictions to \oq safe" fit intervals of the input distributions,
the authors obtain 
\begin{eqnarray}
\amz = 0.1165 &\pm 0.002& {\rm (exp.)} \nonumber \\
&\pm 0.003& {\rm (theo.)}\ ,
\end{eqnarray} 
in NLO QCD. 
Using NNLO QCD calculations wherever available, the same fit gives
\begin{eqnarray}
\amz = 0.1153 &\pm 0.002& {\rm (exp.)} \nonumber \\
&\pm 0.003& {\rm (theo.)}.
\end{eqnarray}
The latter result, however, does not relate to complete NNLO since 
predictions of
jet production cross sections and parts of the DIS structure functions 
are only available in NLO so far.
The previuosly obtained corresponding result, 
$\amz = 0.119 \pm 0.002 \pm 0.003$ \cite{martin2001},
was derived without applying the new \oq safe" fit conditions.

\subsection{$\epem$ Annihilation}

A recent summary of $\as$ determinations from LEP, at the highest
$\epem$ collision energies, and from previous experiments at 
the PETRA and TRISTAN $\epem$ colliders was given in \cite{lepqcd04}.
Apart from the results which were already discussed and presented
in \cite{concise,as2002}, updates on $\as$ from electroweak precision measurements \cite{lepewwg03,lepew-0404}, 
from $\tau$ lepton decays and from a recent combination
of $\as$ determinations from hadronic event shape observables and 
jet rates by the LEP QCD Working Group \cite{lepqcdwg1,lepqcdwg2}.

The most recent combination of the LEP-I and LEP-II 
electroweak precision measurements of all four experiments,
in NNLO QCD, resulted in
\begin{eqnarray}
\amz = 0.1226 &\pm 0.0038& {\rm (exp.)} \nonumber \\
&^{+0.0033}_{-0.0000}& {\rm (M_H)} \nonumber \\
&^{+0.0028}_{-0.0005}& {\rm (QCD)} 
\end{eqnarray}
from $\rz = \Gamma_{had} / \Gamma_{\ell} = 20.767 \pm 0.025$, 
whereby the second error accounts for variations
of the unknown Higgs boson mass between 100 and 900 GeV/c$^2$. 
The third error 
comes from a parametrisation of the unknown higher order QCD corrections,
i.e. from variations of the QCD renormalisation scale 
and renormalistion scheme, see e.g. \cite{concise}.

\begin{figure}[htb]
\begin{center}
\includegraphics[width=73mm, keepaspectratio]{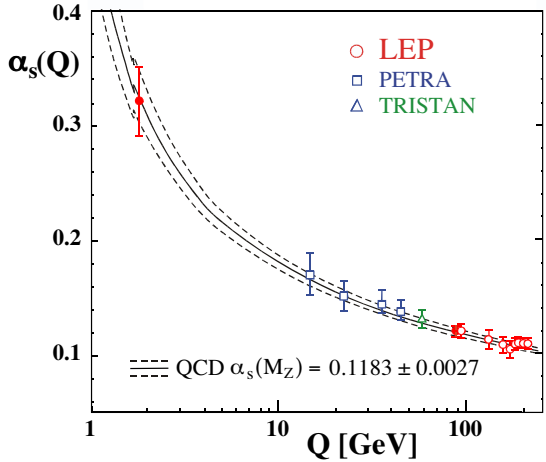}
\end{center}
\vskip-10mm
\caption{\label{asq-LEP}
Summary of measurements of $\asq$ from electroweak precision measurements 
(filled symbols; in NNLO QCD) and from jet and event shape observables
(open symbols; in resummed NLO QCD) in $\epem$ annihilation. 
The dashed curves represent the world average of $\amz$ \cite{lepqcd04}.
}
\end{figure}

In the same analysis \cite{lepew-0404}, 
the fitted leptonic pole cross section,
$\sigma_{\ell}^0  = (2.0003 \pm 0.0027)$~pb, resulted in
\begin{eqnarray}
\amz = 0.1183 &\pm 0.0030& {\rm (exp.)} \nonumber \\
&^{+0.0026}_{-0.0000}& {\rm (M_H)}.
\end{eqnarray}
Since $\sigma_{\ell}^0  = {{12 \pi}\over{M_Z^2}} {{\Gamma_{\ell}^2}
\over {\Gamma_Z^2}}$ and $\Gamma_Z \sim \Gamma_{had}$,
$\sigma_{\ell}$ has a steeper dependence on $\as$ than has $\Gamma_{had}$: in next-to-leading order, 
the QCD coefficient $C_1$ for 
$\Gamma_{had} \sim \left( 1 + \Sigma_n ( C_n \as^n )\right)$, 
$n = 1, 2, 3, ...$, turns to
$2 C_1$ for $\sigma_{\ell}$, $C_2$ turns to $(2 C_2 + C_1^2)$ etc.
The experimental error of $\as$ from $\sigma_{\ell}$ is thus smaller than
that from $\Gamma_{had}$. 
However, with increased QCD-coefficients $C_i$, the renormalisation
scale uncertainty also increases, c.f. 
Equation~13 of \cite{concise},
such that the QCD uncertainty on $\as$ from $\sigma_{\ell}$
is expected to increase w.r.t. $\as$ from $\rz$.

A global fit of all LEP data to determine
$\as$ together with the masses of the $\z0$ boson, 
of the top-quark and of the Higgs boson, gives \cite{lepew-0404}
\begin{equation}
\amz = 0.1200  ^{+0.0031}_{-0.0029} {\rm (exp.)}.
\end{equation}
The latter result is the most precise available from 
combined electroweak fits of the LEP data.

Finally, using all available data from LEP, from SLC and from the Tevatron
(i.e. including direct measurements of the masses of the top-quark and
of the W-boson) results in
\begin{equation}
\amz = 0.1186 \pm 0.0027\ {\rm (exp.)}.
\end{equation}
For the latter two results,
there is no additional uncertainty due to the unknown Higgs mass.
The QCD uncertainties for these particular results of $\as$, however, 
were never determined, and prove to be difficult to be guessed due
to the unknown size of the effective QCD coefficients
that enter the overall fit.
Similar as argued in the case of $\sigma_{\ell}^0$,
the QCD uncertainty on $\Gamma_{had}$ may be a good approximative 
estimate, but cannot simply be applied to 
other observables - especially if they are to be regarded as precision results.

The combined result of $\as$ from $\tau$-decays, in NNLO QCD, 
is \cite{lepqcd04}
\begin{eqnarray}
\as (M_{\tau}) = 0.322 &\pm 0.005& {\rm (exp.)} \nonumber \\
&\pm 0.030& {\rm (theo.)}. 
\end{eqnarray}
When extrapolated to the energy scale $\mz$, this results
in $\amz = 0.1180 \pm 0.0005 {\rm (exp.)} \pm 0.0030 {\rm (theo.)}$.

The overall combination of all LEP results on hadronic event shapes and jet production rates \cite{lepqcdwg2}, using resummed NLO QCD predictions, results in
\begin{eqnarray}
\amz = 0.1202 &\pm 0.0003& {\rm (stat.)} \nonumber \\
&\pm 0.0049& {\rm (syst.)}.
\end{eqnarray}
This analysis also provides precise values of $\as$ in the c.m. energy range of LEP-I and LEP-II, from 91.2 to 206 GeV.
These results, together with those from electroweak precision measurements 
and from $\tau$ decays as given above and also including $\as$ determinations from lower energy $\epem$ colliders, are summarised in Figure~2.

\subsection{Hadron Colliders}

Since the last update of this $\as$ review \cite{as2002},
no new measurements of $\as$ from hadron colliders were reported.
In general, the available results from earlier studies, all in NLO QCD, 
are compatible with but not really competetive to those obtained from
DIS and from $\epem$ annihilation.
This is due to the sum of systematic uncertainties, from 
higher order QCD as well as from 
the available Monte Carlo generators, from underlying events and beam
remnants, and from the energy calibration of detectors. 
Necessary improvements, like QCD predictions in NNLO and/or including 
resummation, corrections for nonperturbative effects, improved tools 
like new and more reliable jet algorithms, more data statistics and 
possibly more data at different collision energies are, however, 
underway and should be in place at the startup of the Large Hadron 
Collider.

\section{SUMMARY AND WORLD AVERAGE}

\begin{figure*}[htb]
\begin{center}
\includegraphics[scale=.55]{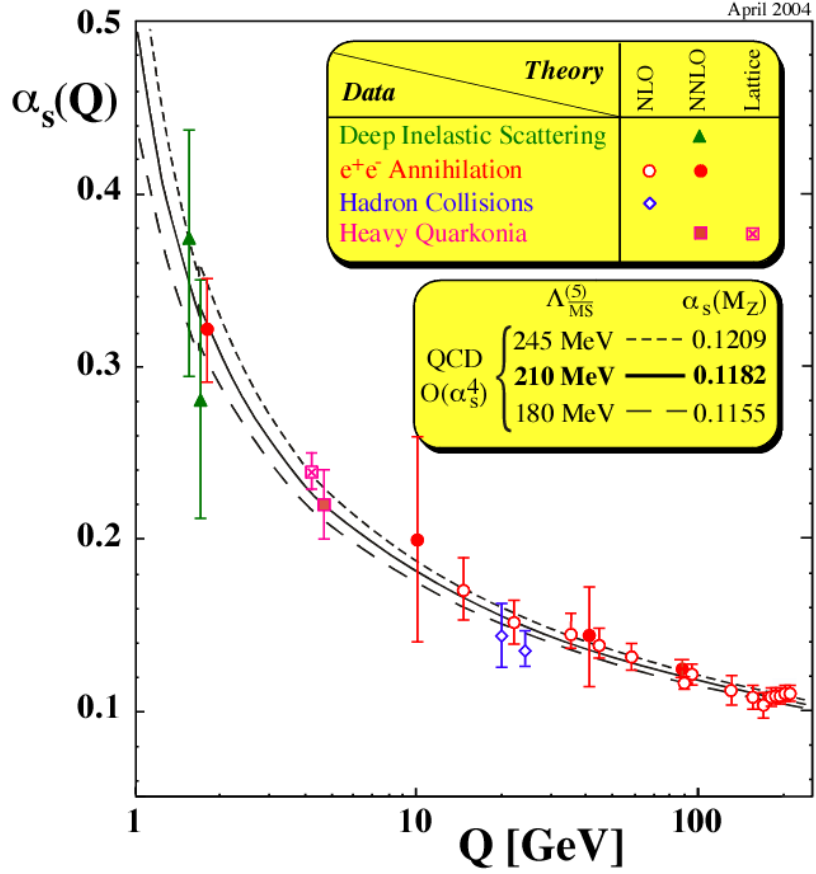}
\end{center}
\vskip-10mm
\caption{\label{asq}
Summary of measurements of $\asq$. 
Results which are based on fits of $\amz$ to data in $ranges$ of $Q$, 
assuming the QCD running of $\as$,
are not shown here but are included
in the overall summary of $\amz$, see Figure~4 and Table~1.}
\end{figure*}

A summary of all significant measurements of $\as$, 
as discussed in \cite{concise,as2002} and
with updates and new measurements presented in this review, 
is given in Table~1.

\begin{figure}[htb]
\begin{center}
\includegraphics[scale=.5]{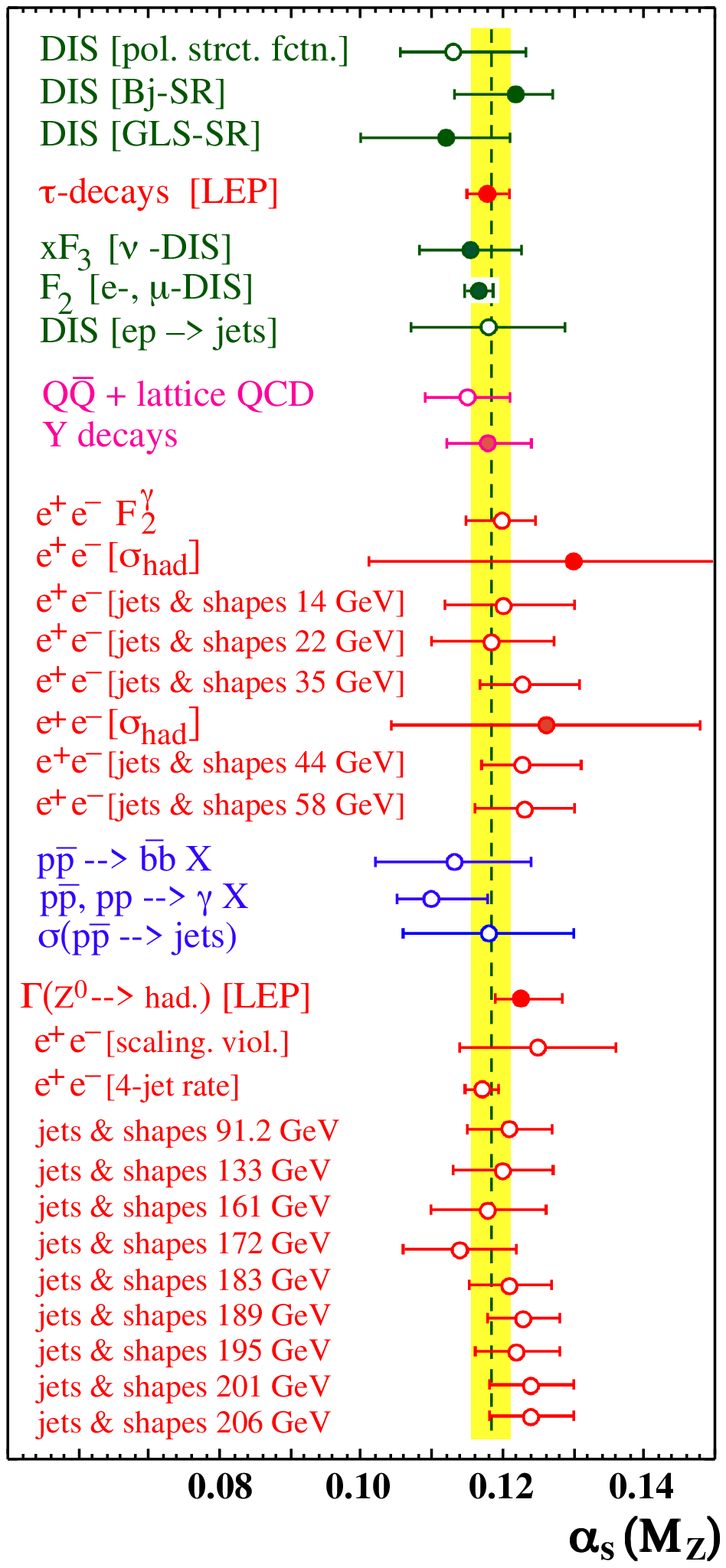}
\end{center}
\vskip-10mm
\caption{\label{asmz}
Summary of measurements of $\amz$. 
Filled symbols represent results based on complete NNLO QCD calculations.}
\end{figure}

The values of $\as (Q)$ are presented in Figure~3, as a 
function of the energy scale $Q$ where the measurement was carried 
out.
The data provide 
significant evidence for the running of $\as$, in good agreement with the 
QCD prediction.

Therefore it is appropriate to extrapolate all results of $\as (Q)$ to a common 
value of energy, which is usually the rest energy of the $\z0$ boson,  $\mz$.
As described in \cite{concise}, the QCD evolution of $\as$ with 
energy, using the full 4-loop expression \cite{4-loop} with 3-loop matching 
\cite{matching} at the pole masses of the charm- and the bottom-quark,
$M_c = 1.7\ GeV$ and $M_b = 4.7\ GeV$, is applied to all results of $\as (Q)$ 
which were obtained at energy scales $Q \ne \mz$.

The corresponding values of $\amz$ are tabulated in the $4^{th}$ column of 
Table~1; column 5 and 6 indicate the  contributions of the experimental and 
the theoretical unceratinties to the overall errors assigned to $\amz$.
All values of $\amz$ are graphically displayed in Figure~4.
Within their individual uncertainties, there is perfect agreement between 
all results. 
This justifies to evaluate an overall world average value, $\wamz$.
As discussed e.g. in \cite{concise}, however, the combination of all these 
results to an overall average, and even more so for the overall
uncertainty to be assigned to this average,
is not trivial due to the supposedly large 
but unknown correlations between invidual results, especially through 
common prejudices and biases within the theoretical calculations.

For combining all or subsets of the results summarised in Table~1 into 
average values of $\amz$, the same procedures as utilised  
in \cite{concise} are being used.
Averages $\wamz$ for all and for subsets of $\as$-results, together 
with the corresponding uncertainties $\dwas$ are summarised in 
Table~\ref{tab:aserr}.
As already discussed in \cite{concise}, the overall uncertainties decrease 
if the averaging process is restricted to those which accomplished a 
minimum precision, i.e. a total error of $\Delta \as \le 0.008$, while the
value of $\wamz$ is almost unaffected by such a restriction - c.f. rows~1 
and~2.

There is a sufficiently large number of redults which is based on complete 
NNLO QCD, such that $\wamz$ can be reliably calculated from this subset 
(see rows 5 to 8 of Table~\ref{tab:aserr}).
Due to the improved completenes of the perturbation series, these results 
are believed to be more reliable and better defined than all the others 
which are complete to (resummed) NLO.

The world average of $\amz$ is finally determined from those NNLO results
that have total errors less than 0.008, namely
\begin{eqnarray}
\rm{DIS\ [Bj-SR]:} &\amz =& 0.121 ^{+0.005}_{-0.009}\ ,\nonumber \\
\rm{\tau\ decays:} &\amz =& 0.1180 \pm 0.0030\ ,\nonumber \\
\rm{DIS\ [\nu, xF_3]:} &\amz =& 0.119 ^{+0.007}_{-0.006}\ ,\nonumber \\
\rm{DIS\ [e/\mu, xF_2]:} &\amz =& 0.1166 \pm 0.0022 ,\nonumber \\
\rm{\Upsilon\ decays:} &\amz =& 0.118 \pm 0.006\ ,\nonumber \\
{\rm \Gamma (Z \rightarrow had):} &\amz =& 0.1226 ^{+0.0058}_{-0.0038} \ .\nonumber
\end{eqnarray}
For defining the overall total uncertainty, an \oq optimized
correlation" error is calculated from the error covariance matrix, assuming
an overall correlation factor between the total errors of all measurements.
This factor is adjusted such that the overall $\chi^2$ equals one per degree
of freedom.
This results in the new world average of
\begin{equation}
\wamz = 0.1182 \pm 0.0027\ ,
\end{equation}
c.f. row~6 of Table~\ref{tab:aserr},
which is practically identical to the final result of the 
previous summary \cite{as2002}, $\wamz = 0.1183 \pm 0.0027$.

The choice of results which contribute to the determination of the world average
value of $\wamz$ is, similar as the calculation and definition of its
overall uncertainty, arbitrary to a large degree.
A matter of discussion sometimes is to include the overall fit result to the
combined LEP electroweak precision data (Equation~8) instead of, as it is done 
here, using the result from $R_Z$ (Equation~6), because the overall fit uses more 
information,  the experimental error of $\as$ is smaller, and
there is no uncertainty on ${\rm M_H}$ in this case.
With this replacement, and assuming the same QCD uncertainty
as for $\as$ from $R_Z$ (which may not be appropriate, see the discussion in 
Section~2.2), results in
$\wamz = 0.1180 \pm 0.0029$.
Replacing the result from Equation~6 with the one from Equation~9, i.e. with the world fit of electroweak data, gives
$\wamz = 0.1177 \pm 0.0031$.
In both these cases, $\wamz$ slightly decreases and the overall uncertainty 
increases, but fully within the assigned error.
The slight increase of $\dwas$ is due to an increased correlation factor of 0.80 and 0.89, respectively, which is assumed to assure that the overall $\chi^2$ is
unity per degree of freedom.
For reasons of clarity and of strict definitions of uncertainties, the
usage of the result from $R_Z$ (Equation~6) is preferred.

The world average of $\wamz = 0.1182 \pm 0.0027$ corresponds to the following 
values of the QCD scale $\lamsb$ for different numbers of 
quark flavours $N_f$, 
evaluated using the full 4-loop expansion
of $\as$ and 3-loop matching at the quark thresholds (c.f. Equations~4 
and~8 of \cite{concise}):
\begin{eqnarray}
\lamsb^{N_f=5} &=& 210^{+34}_{-30}\ {\rm MeV}  \\
\lamsb^{N_f=4} &=& 294^{+41}_{-39}\ {\rm MeV}\ .  
\end{eqnarray}

\renewcommand{\arraystretch}{1.2}
\begin{table*}[h]
{
\caption{
World summary of measurements of $\as$ (status of April 2004):
DIS = deep inelastic scattering; GLS-SR = Gross-Llewellyn-Smith sum rule;
Bj-SR = Bjorken sum rule;
(N)NLO = (next-to-)next-to-leading order perturbation theory;
LGT = lattice gauge theory;
resum. = resummed NLO. 
\label{tab:astab}}
\begin{center}
\begin{tabular}{|l|c|c|c|c c|c|c|}
   \hline 
  & Q & & &  \multicolumn{2}{c|}
{$\Delta \amz $} & & \\ 
Process & [GeV] & $\alpha_s(Q)$ &
  $ \amz$ & exp. & theor. & Theory & refs.\\
\hline \hline 
DIS [pol. SF] & 0.7 - 8 & & $0.113\ ^{+\ 0.010}
  _{-\ 0.008}$ & $\pm 0.004$ & $^{+0.009}_{-0.006}$ & NLO &
\cite{bluemlein02}\\
DIS [Bj-SR] & 1.58
  & $0.375\ ^{+\ 0.062}_{-\ 0.081}$ & $0.121\ ^{+\ 0.005}_{-\ 0.009}$ & 
  -- & -- & NNLO & \cite{bjsr}\\
DIS [GLS-SR] & 1.73
  & $0.280\ ^{+\ 0.070}_{-\ 0.068}$ & $0.112\ ^{+\ 0.009}_{-\ 0.012}$ & 
  $^{+0.008}_{-0.010}$ & $0.005$ & NNLO & \cite{gls-recent}\\
$\tau$-decays 
  & 1.78 & $0.322 \pm 0.030$ & $0.1180 \pm 0.0030$
  & 0.0005 &  0.0030 & NNLO & \cite{lepqcd04}\\
DIS [$\nu$; ${\rm x F_3}$]  & 2.8 - 11
  & 
   & $0.119\ ^{+\ 0.007}_{-\ 0.006}$   &
    $ 0.005 $ & $^{+0.005}_{-0.003}$ & NNLO & \cite{kataev2001}\\
DIS [e/$\mu$; ${\rm F_2}$]
     & 1.9 - 15.2 &      & $0.1166 \pm 0.0022$ & $ 0.0009$ &
     $ 0.0020$ & NNLO & \cite{yndurain2001,as2002}\\
DIS [e-p $\rightarrow$ jets]
     & 6 - 100 &  & $0.120 \pm 0.005$ & $ 0.002$ &
     $0.004 $ & NLO & \cite{here}\\
${\rm Q\overline{Q}}$ states
     & 4.1 & $0.239\ ^{+\ 0.012}_{-\ 0.010}$ & $0.121 \pm 0.003 
     $ & 0.000 & 0.003
     & LGT & \cite{lgt-3}\\
$\Upsilon$ decays
     & 4.75 & $0.217 \pm 0.021$ & $0.118 \pm 0.006
     $ & -- & -- & NNLO & \cite{Ydec-3rd}\\
$\epem$ [${\rm F^{\gamma}_2}$]
     & 1.4 - 28 &  & $0.1198\ ^{+\ 0.0044}_{-\ 0.0054}$ 
     & 0.0028 & $^{+\ 0.0034}_{-\ 0.0046}$ & NLO & \cite{lep-2gamma}\\
$\epem$ [$\sigma_{\rm had}$] 
     & 10.52 & $0.20\ \pm 0.06 $ & $0.130\ ^{+\ 0.021\ }_{-\ 0.029\ }$
     & $\ ^{+\ 0.021\ }_{-\ 0.029\ }$ & 0.002 & NNLO & \cite{cleo-rhad}\\
$\epem$ [jets \& shps]  & 14.0 & $0.170\ ^{+\ 0.021}_{-\ 0.017}$ &
   $0.120\ ^{+\ 0.010}_{-\ 0.008}$ &  0.002 & $^{+0.009}_{-0.008}$
   & resum & \cite{fernandez-2002}\\
$\epem$ [jets \& shps]  & 22.0 & $0.151\ ^{+\ 0.015}_{-\ 0.013}$ &
   $0.118\ ^{+\ 0.009}_{-\ 0.008}$ &  0.003 & $^{+0.009}_{-0.007}$
   & resum  & \cite{fernandez-2002}\\
$\epem$ [jets \& shps] & 35.0 & $ 0.145\ ^{+\ 0.012}_{-\ 0.007}$ &
   $0.123\ ^{+\ 0.008}_{-\ 0.006}$ &  0.002 & $^{+0.008}_{-0.005}$
   & resum  & \cite{fernandez-2002}\\
$\epem$ [$\sigma_{\rm had}$]  & 42.4 &
 $0.144 \pm 0.029$ &
   $0.126 \pm 0.022$ & $0.022
   $ & 0.002 & NNLO & \cite{haidt,concise}\\
$\epem$ [jets \& shps] & 44.0 & $ 0.139\ ^{+\ 0.011}_{-\ 0.008}$ &
   $0.123\ ^{+\ 0.008}_{-\ 0.006}$ & 0.003 & $^{+0.007}_{-0.005}$
   & resum  & \cite{fernandez-2002}\\
$\epem$ [jets \& shps]  & 58.0 & $0.132\pm 0.008$ &
   $0.123 \pm 0.007$ & 0.003 & 0.007 & resum & \cite{as-tristan}\\
$\p\bar{\p} \rightarrow {\rm b\bar{b}X}$
    & 20.0 & $0.145\ ^{+\ 0.018\ }_{-\ 0.019\ }$ & $0.113 \pm 0.011$ 
    & $^{+\ 0.007}_{-\ 0.006}$ & $^{+\ 0.008}_{-\ 0.009}$ & NLO 
    & \cite{ua1-bb}\\
${\rm p\bar{p},\ pp \rightarrow \gamma X}$  & 24.3 & $0.135
 \ ^{+\ 0.012}_{-\ 0.008}$ &
  $0.110\ ^{+\ 0.008\ }_{-\ 0.005\ }$ & 0.004 &
  $^{+\ 0.007}_{-\ 0.003}$ & NLO & \cite{ppgam-recent}\\
${\sigma (\rm p\bar{p} \rightarrow\  jets)}$  & 40 - 250 &  &
  $0.118\pm 0.012$ & $^{+\ 0.008}_{-\ 0.010}$ & $^{+\ 0.009}_{-\ 0.008}$ & 
  NLO & \cite{cdf-jet}\\
$\epem$ [$\Gamma (\rm{Z \rightarrow had})$]
    & 91.2 & $0.1226^{+\ 0.0058}_{-\ 0.0038}$ & 
    $0.1226^{+\ 0.0058}_{-\ 0.0038}$ &
   $\pm 0.0038$ & $^{+0.0043}_{-0.0005}$ & NNLO & \cite{lepew-0404}\\
$\epem$ scal. viol. & 14 - 91.2 &  & $0.125 \pm 0.011$ & 
   $^{+\ 0.006}_{-\ 0.007}$ & 0.009 & NLO & \cite{concise}\\
$\epem$ 4-jet rate &  91.2 & $0.1170 \pm 0.0026$ & $0.1170 \pm 0.0026$ & 
  0.0001 & 0.0026 & NLO & \cite{a-4j}\\
$\epem$ [jets \& shps] &
    91.2 & $0.121 \pm 0.006$ & $0.121 \pm 0.006$ & $ 0.001$ & $
0.006$ & resum & \cite{concise}\\
$\epem$ [jets \& shps]  & 133 & $0.113\pm 0.008$ &
   $0.120 \pm 0.007$ & 0.003 & 0.006 & resum & \cite{concise}\\
$\epem$ [jets \& shps]  & 161 & $0.109\pm 0.007$ &
   $0.118 \pm 0.008$ & 0.005 & 0.006 & resum & \cite{concise}\\
$\epem$ [jets \& shps]  & 172 & $0.104\pm 0.007$ &
   $0.114 \pm 0.008$ & 0.005 & 0.006 & resum & \cite{concise}\\
$\epem$ [jets \& shps]  & 183 & $0.109\pm 0.005$ &
   $0.121 \pm 0.006$ & 0.002 & 0.005 & resum & \cite{concise}\\
$\epem$ [jets \& shps] & 189 & $0.109\pm 0.004$ &
   $0.121 \pm 0.005$ & 0.001 & 0.005 & resum & \cite{concise}\\
$\epem$ [jets \& shps] & 195 & $0.109\pm 0.005$ &
   $ 0.122\pm 0.006$ & 0.001 & 0.006 & resum & \cite{as2002}\\
$\epem$ [jets \& shps] & 201 & $0.110\pm 0.005$ &
   $ 0.124\pm 0.006$ & 0.002 & 0.006 & resum & \cite{as2002}\\
$\epem$ [jets \& shps] & 206 & $0.110\pm 0.005$ &
   $ 0.124\pm 0.006$ & 0.001 & 0.006 & resum & \cite{as2002}\\
\hline
\end{tabular}
\end{center}
}
\end{table*}

\renewcommand{\arraystretch}{1.2}
\begin{table*}[htb]
\caption{
Average values of $\wamz$ and averaged uncertainties, for several subsamples 
of the available data. 
The result printed in bold-face is taken as the new
world average value of $\wamz$.
\oq $- F_2$" denotes omission of the result from  
$F_2$ structure function fits \cite{yndurain2001}.
\label{tab:aserr} }
\begin{center}
  {
\begin{tabular}{|c|l|c|c|c||c|}
   \hline
& & & opt. corr. & overall & uncorrel. \\
row & sample \hfill (entries)& $\wamz$ & $\dwas$ & correl. &
  $\dwas$  \\
\hline
1 & all \hfill (32)     & 0.1190 & 0.0038 & 0.69 & 0.0009 \\
2 &\ "\ $\Delta \as\le 0.008$ \hfill (23)
                        & 0.1192 & 0.0034 & 0.63 & 0.0010 \\
3 & all - $F_2$ \hfill (31)
                        & 0.1196 & 0.0043 & 0.71 & 0.0010 \\
4 &\ "\ $\Delta \as\le 0.008$\hfill (22)
                        & 0.1198 & 0.0039 & 0.68 & 0.0010 \\
 & & & & & \\
5 & NNLO only \hfill (9) & 0.1182 & 0.0031 & 0.68 & 0.0015 \\
6 &\ "\ $\Delta \as\le 0.008$ \hfill (6)
                        &\bf  0.1182 & \bf 0.0027 & 0.61 
                        & 0.0015 \\
7 & NNLO - $F_2$ \hfill (8)
                        & 0.1195 & 0.0042 & 0.75 & 0.0019 \\
8 &\ "\ $\Delta \as\le 0.008$ \hfill (5)
                        & 0.1196 & 0.0038 & 0.74 & 0.0020 \\
&  & & & & \\
9 & NLO only \hfill (23)
                        & 0.1197 & 0.0044 & 0.69 & 0.0011 \\
10 &\ "\ $\Delta \as\le 0.008$\hfill (17)
                        & 0.1199 & 0.0040 & 0.65 & 0.0012 \\

&  & & & & \\
11 & DIS only \hfill (6) & 0.1169 & 0.0033 & 0.76 & 0.0018 \\
12 & DIS - $F_2$\hfill (5) & 0.1185 & 0.0060 & 0.77 & 0.0031 \\
13 & $\epem$ only\hfill (22) & 0.1199 & 0.0044 & 0.77 & 0.0012 \\
\hline
\end{tabular} }
\end{center} 
\end{table*}

%


\begin{thebibliography}{99}
%
\bibitem{concise} 
S. Bethke, J. Phys. G26 (2000) R27; hep-ex/0004021.
%
\bibitem{as2002} 
S. Bethke, Proc. of the {\it QCD 02 High-Energy Physics International
Conference in QCD}, Montpellier (France) July 2-9, 2002; 
hep-ex/0211012.
%
\bibitem{zeus03}
ZEUS Collaboration, S. Chekanov et al., Phys. Lett. B560 (2003) 7,
hep-ex/0212064.
%
\bibitem{buschhorn}
G. Buschhorn, hep-ex/0406038.
%
\bibitem{bluemlein02}
J. Bl\"umlein and H. B\"ottcher, Nucl. Phys. B636 (2002) 225, hep-ph/0203155.
%
\bibitem{mrst03}
A.D. Martin, R.G. Roberts, W.J. Stirling and R.S. Thorne,
hep-th/0308087.
%
\bibitem{martin2001}
A.D. Martin et al., Eur. Phys. J. C23 (2002) 73, hep-ph/0110215.
%
\bibitem{lepqcd04}
S. Bethke, hep-ex/0406058.
%
\bibitem{lepewwg03}
The LEP Electroweak Working Group and the LEP experiments 
ALEPH, DELPHI, L3 and OPAL, hep-ex/0312023.
%
\bibitem{lepew-0404}
The LEP Collaborations ALEPH, DELPHI, L3 and OPAL; combined results presented 
at the 2004 Winter Conferences (April 2004), \\ 
http://lepewwg.web.cern.ch/LEPEWWG /stanmod/winter2004\_results.
%
\bibitem{lepqcdwg1}
The LEP QCD Working Group, R.W.L. Jones et al.,
JHEP 0312 (2003) 007, hep-ph/0312016.
%
\bibitem{lepqcdwg2}
The LEP QCD Working Group, S. Banerjee et al.,
in preparation to be published.
%
\bibitem{bjsr}
J. Ellis and M. Karliner, Phys. Lett B341 (1995) 397.
%
\bibitem{gls-recent}
CCFR Collaboration,
J.H. Kim et al., Phys. Rev. Lett. 81 (1998) 3595.
%
\bibitem{kataev2001}
A.L. Kataev et al., hep-ph/0106221.
%
\bibitem{yndurain2001}
J. Santiago, F.J. Yndurain, hep-ph/0102247; Nucl. Phys. B611 (2001) 447.
%
\bibitem{here}
{\it this review}
%
\bibitem{lgt-3}
C. Davies et al., Nucl. Phys. Proc. Suppl. 119 (2003) 595, hep-lat/0209122.
%
\bibitem{Ydec-3rd}
A. Penin, A.A. Pivovarov, Phys. Lett. B435 (1998) 413.
%
\bibitem{lep-2gamma}
S. Albino et al., hep-ph/0205069; Phys. Rev. Lett. 89 (2002) 122004.
%
\bibitem{cleo-rhad}
CLEO Collaboration, R. Ammar et al., Phys. Rev. D57 (1998) 1350.
%
\bibitem{fernandez-2002}
P.A. Movilla Fernandez, hep-ex/0205014.
%
\bibitem{haidt}
D. Haidt, in {\it Directions in High Energy
Physics} Vol 14, Precision Tests of the Standard
Electroweak Model, ed. P. Langacker, World Scientific, 1995.
%
\bibitem{as-tristan}
TOPAZ Collaboration, Y. Ohnishi et al. Phys. Lett. B 313 (1993) 475.
%
\bibitem{ua1-bb}
UA1 Collaboration, C. Albajar et al., Phys. Lett. B369 (1996) 46.
%
\bibitem{ppgam-recent}
UA6 Collaboration, M. Werlen et al., Phys. Lett. B452 (1999) 201.
%
\bibitem{cdf-jet}
T. Affolder et al., CDF collaboration, hep-ex/0108034; 
Phys. Rev. Lett. 88 (2002) 042001.
%
\bibitem{a-4j}
A. Heister et al., ALEPH Collaboration, Eur. Phys. J. C27 (2003) 1.
%
\bibitem{4-loop}
T. van Ritbergen et al., Phys. Lett. B400 (1997) 379.
%
\bibitem{matching}
K.G. Chetyrkin et al., Phys. Rev. Lett. 79 (1997) 2184.
%


\end{thebibliography}
\end{document}